\documentclass[preprint,prb]{revtex4}%
\usepackage{amsfonts}
\usepackage{amsmath}
\usepackage{amssymb}
\usepackage{graphicx}%
\providecommand{\U}[1]{\protect\rule{.1in}{.1in}}
\begin{document}
\title{Quantum interference and spin polarization on Rashba double quantum dots}
\author{Kuo-Wei Chen and Ching-Ray Chang}
\affiliation{Department of Physics, National Taiwan University, Taipei 10617, Taiwan}

\begin{abstract}
We report on the quantum interference and spin accumulation on double quantum
dots with Rashba spin-orbit coupling and electron-electron interaction, based
on the Keldysh nonequilibrium Green function formalism. It is shown that
Rashba spin-orbit interaction can strongly affect the conductance spectrum. By
gradually increasing the Rashba parameter from zero, Fano resonances in strong
overlap regime continuously evolve to resolve antiresonances. This transition
is ascribed to the phase shift of couplings between molecular states and the
lead due to spin precession. We also show that both bias and Rashba effect
strengthen the induced spin polarization in this device. For particular energy
position, up- and down-spin electron occupation can intersect to form a
crossing point. Spin polarization on different side of this point has opposite
sign in consequence. The magnitude and direction of spin polarization are
therefore controllable by tuning the dot levels and the Rashba parameter
through gates.

\end{abstract}
\maketitle

\section{Introduction}

Spin-related phenomena in semiconductor nanostructures are important subjects
and of fundamental interest in condensed matter physics and material
science.\cite{Engel04} In particular, studying the spin-orbit interaction,
combined with applied external fields, to manipulate the electron spin and to
generate the spin-polarized current in spintronic devices is strongly
desired.\cite{Sun08} Among different structures a great deal of experimental
and theoretical researches had been reported on quantum transport properties
through quantum dot (QD) systems due to potential applications, from
spintronic devices to quantum information.\cite{Loss98} Control of the spin
states of electrons confined in QDs is the challenges to build these
spin-based devices.

Mechanisms inducing spin-orbit coupling in QD systems, including both
structure and bulk inversion asymmetries, have been recently
studied,\cite{Lopez07,Golovach08} in the hope that spin states of electrons
can be coherently controlled.\cite{Sun06,Lobos08,Alves08} Among these two
mechanisms the Rashba spin-orbit (RSO) coupling is of more potential
applicability, because of its coupling strength tunability via the interface
electric field controllable by either applied gate voltage or
doping.\cite{Nitta97} A built-in asymmetric potential underlying Rashba
mechanism causes a moving electron to precess about the effective magnetic
field. Throughout the transport a precession angle of an electron spin is thus
generated. Quantum interference consequent on the electron spins with
different precession angles is modulated by tunable RSO coupling strength, as expected.

A variety of coupled QDs structures\ with sizes smaller than electron
coherence length have been achieved in mesoscopic solid-state
circuits.\cite{Rohrlich07} The important characteristics of preserved
coherence of electrons in QD systems, such as the Aharonov-Bohm
oscillation\cite{Yacoby95,Sigrist07} and\ the Fano
effect\cite{Gores00,Kobayashi02,Johnson04}, have been widely observed. The
latter arises from the interference between a discrete state and the
continuum.\cite{Fano61} For coupled double QDs (DQD), an interdot coupling
brings coherent hopping through tunneling barrier between two dots, which
causes the level repulsion of DQD and meanwhile complicates the electron
transport through the system.\cite{Joe05} On the way to understand the
electron transport considering RSO interaction, the dephasing and interference
can be studied via the Fano effect, which had been proved to be an effective
method.\cite{Clerk01} On the other hand the behavior of spins in dots each and
the relation in between, corresponding to tunable RSO effect, are also required.

In this paper we study the quantum transport through a ring with serially
coupled DQD. Quantum interference arises from the electron waves passing
through different paths under the influence of the RSO interaction. For
electrons in one lead, the ways across the system to the other contain a
bridge channel and four molecular states separated by the interdot coupling
and the electron-electron interaction. Concerning the system out of
equilibrium associated with a finite bias, the spin-dependent conductance is
calculated based on the Keldysh Green function formalism.\cite{Keldysh65} We
show that the RSO interaction strongly affects the resonances in conductance
both in weak and strong bias condition. When varying the Rashba parameter, the
phases of tunneling couplings between molecular states and the lead are
gradually changed, accompanied with the formations of Fano antiresonances. We
also study the spin accumulation phenomenon on DQD. Above the Kondo
temperature the dots are unpolarized without external magnetic field, while
spin-polarized occupation is shown to exist in this device due to the combined
effect of a bridge channel and the RSO coupling. Advantageously the spin
polarization (SP) is sensitive to the level position and the Rashba coupling
strength on dots. The manipulation of the SP, such as magnitude and direction,
can therefore be easily achieved by tuning gate voltage, which may be useful
in the design of spintronic devices.

The paper is organized as follows. In Sec. II we describe the model and the
self-consistent procedure used to obtain the nonequilibrium Green functions in
current expression. In Sec. III we show the interference phenomena both in
weak and strong bias condition. Differential conductance is discussed in three
cases of different controlling ways of the energy levels on two dots. In Sec.
IV we show the spin accumulation induced in this device and study the behavior
of SP as the bias and the RSO coupling are varied. Then final conclusions are
gathered in Sec. V.

\section{The model}%

%TCIMACRO{\FRAME{fhFU}{1.9484in}{1.9052in}{0pt}{\Qcb{(Color online) Double
%quantum dots coupled serially to leads, denoted as L and R, with Rashba
%spin-orbit coupling and on-site Coulomb repulsion.}}{}{fig1.ps}%
%{\special{ language "Scientific Word";  type "GRAPHIC";
%maintain-aspect-ratio TRUE;  display "USEDEF";  valid_file "F";
%width 1.9484in;  height 1.9052in;  depth 0pt;  original-width 5.0557in;
%original-height 4.9398in;  cropleft "0";  croptop "1";  cropright "1";
%cropbottom "0";  filename 'FIG1.ps';file-properties "XNPEU";}}}%
%BeginExpansion
\begin{figure}
[h]
\begin{center}
\includegraphics[
height=1.9052in,
width=1.9484in
]%
{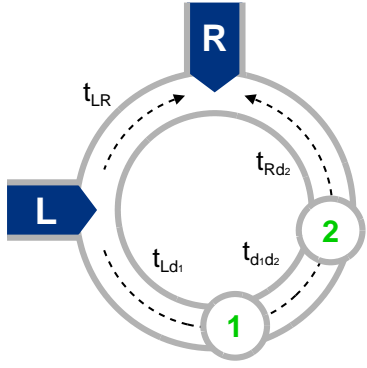}%
\caption{(Color online) Double quantum dots coupled serially to leads, denoted
as L and R, with Rashba spin-orbit coupling and on-site Coulomb repulsion.}%
\end{center}
\end{figure}
%EndExpansion
The device is illustrated in Fig. 1, where the ring-dot structure is realized
on the heterostructure confining a two-dimensional electron gas (2DEG). In
semiconductor the energy separation between single-electron levels has been
reported to be a considerable energy scale due to the small effective mass.
Therefore, only single energy level near the Fermi surface is assumed to be
relevant in each dot. Two dots, with both RSO coupling and Coulomb interaction
taken into account, are connected in series and embedded in one arm of the
ring, and then attached to two normal metal leads. No interaction exists on
the bridge channel. Experimentally one can tune the electron density on the
dot and the couplings to the leads by using several gates.\cite{Fuhrer06}

We begin with the following model Hamiltonian%
\begin{equation}
H=H_{0}+H_{B}+H_{D}+H_{T},
\end{equation}
where the terms%
\begin{equation}
H_{0}=\sum\limits_{\alpha\mathbf{k}\sigma}\epsilon_{\alpha\mathbf{k}}%
n_{\alpha\mathbf{k}\sigma}+\sum\limits_{i\sigma}\left(  \epsilon_{i}%
n_{i\sigma}+Un_{i\uparrow}n_{i\downarrow}\right)  ,
\end{equation}%
\begin{equation}
H_{B}=\sum\limits_{\mathbf{kk}^{\prime}\sigma}t_{LR}\left(  a_{L\mathbf{k}%
\sigma}^{\dag}a_{R\mathbf{k}^{\prime}\sigma}+a_{R\mathbf{k}^{\prime}\sigma
}^{\dag}a_{L\mathbf{k}\sigma}\right)  ,
\end{equation}%
\begin{equation}
H_{D}=\sum\limits_{\sigma}t_{d_{1}d_{2}}\left(  d_{1\sigma}^{\dag}d_{2\sigma
}+d_{2\sigma}^{\dag}d_{1\sigma}\right)  ,
\end{equation}
and%
\begin{equation}
H_{T}=\sum\limits_{\mathbf{k}\sigma}t_{Ld_{1}}a_{L\mathbf{k}\sigma}^{\dag
}d_{1\sigma}+t_{Rd_{2}}e^{-i\sigma\varphi}a_{R\mathbf{k}\sigma}^{\dag
}d_{2\sigma}+H.c.
\end{equation}
are explained below. The number operators for the corresponding states of the
dot spin and the lead spin are given by $n_{i\sigma}=d_{i\sigma}^{\dag
}d_{i\sigma}$ and $n_{\alpha\mathbf{k}\sigma}=a_{\alpha\mathbf{k}\sigma}%
^{\dag}a_{\alpha\mathbf{k}\sigma}$, respectively, where the fermionic operator
$d_{i\sigma}$ ($d_{i\sigma}^{\dag}$) destroys (creates) an electron with spin
$\sigma=\uparrow,\downarrow$ on the $i$th ($i=1,2$) dot, and $a_{\alpha
\mathbf{k}\sigma}$ ($a_{\alpha\mathbf{k}\sigma}^{\dag}$) destroys (creates) an
electron with energy $\epsilon_{\alpha\mathbf{k}}$ in the lead $\alpha=L,R$.

The first term in $H_{0}$ describes the leads, which are modeled by Fermi sea
with energy $\epsilon_{\alpha\mathbf{k}}$. Each lead is filled up to an
electrochemical potential $\mu_{\alpha}$, and the occupation number obeys the
Fermi distribution $f_{\alpha}\left(  \omega\right)  =\left\{  \exp\left[
\left(  \omega-\mu_{\alpha}\right)  /k_{B}T\right]  +1\right\}  ^{-1}$. The
second and third terms in $H_{0}$ correspond to the isolated dots. Each dot
consists of a single level $\epsilon_{i}$ and an on-site Coulomb repulsion
with constant strength $U$. Note that the dot levels are assumed to be spin
degenerate since the effect of the RSO coupling simply contributes to an extra
phase in the tunneling matrix elements, when choosing a space-dependent spin
coordinate.\cite{Sun05} The terms $H_{B}$ and $H_{D}$\ account for the bridge
channel between two leads with coupling strength $t_{LR}$ and the hopping
between two dots with coupling strength $t_{d_{1}d_{2}}$, respectively. The
last term $H_{T}$ is the couplings of the dots to the leads. For simplicity we
have assumed all the tunneling matrix elements to be spin-independent. Owing
to the RSO coupling there is a spin-dependent phase $\varphi=\alpha_{R}%
m^{\ast}L/\hbar^{2}$ generated in the path through the dots, with $\alpha_{R}$
being the Rashba parameter and $L$ being the size of the DQD.

Applying the same procedure introduced in Ref. \onlinecite{Sun05}, we now
analyze the quantum transport property of this device through the Keldysh
nonequilibrium Green function technique. The charge current flowing from the
left lead into the ring can be calculated from the time evolution of the
occupation number for electrons in the left lead, which is written
as\cite{Meir92}%
\begin{equation}
I_{\sigma}=\frac{2e}{\hbar}\int\frac{d\omega}{2\pi}\operatorname{Re}\left[
t_{Ld_{1}}G_{d_{1}L\sigma}^{<}\left(  \omega\right)  +t_{LR}G_{RL\sigma}%
^{<}\left(  \omega\right)  \right]  ,
\end{equation}
where the lesser Green function $G_{ij\sigma}^{<}\left(  \omega\right)  $
correlates the states in $i$ and $j$ with spin $\sigma$. The spin-dependent
conductance is thus defined as $G_{\sigma}=dI_{\sigma}/dV$. From the kinetic
equation and the assumption of ideal leads, the lesser Green functions are
related to the retarded and advanced Green functions through%
\begin{equation}
G_{\sigma}^{<}=G_{\sigma}^{r}g_{\sigma}^{r-1}g_{\sigma}^{<}g_{\sigma}%
^{a-1}G_{\sigma}^{a}.
\end{equation}
In order to obtain the lesser Green functions, the retarded Green functions
are calculated by the Dyson equation%
\begin{equation}
G_{\sigma}^{r}=g_{\sigma}^{r}+g_{\sigma}^{r}\Sigma_{\sigma}^{r}G_{\sigma}^{r},
\end{equation}
where the retarded Green function $G_{\sigma}^{r}$ in the local basis is a
$4\times4$ matrix%
\begin{equation}
G_{\sigma}^{r}\equiv\left[
\begin{array}
[c]{cccc}%
G_{LL\sigma}^{r} & G_{LR\sigma}^{r} & G_{Ld_{1}\sigma}^{r} & G_{Ld_{2}\sigma
}^{r}\\
G_{RL\sigma}^{r} & G_{RR\sigma}^{r} & G_{Rd_{1}\sigma}^{r} & G_{Rd_{2}\sigma
}^{r}\\
G_{d_{1}L\sigma}^{r} & G_{d_{1}R\sigma}^{r} & G_{d_{1}d_{1}\sigma}^{r} &
G_{d_{1}d_{2}\sigma}^{r}\\
G_{d_{2}L\sigma}^{r} & G_{d_{2}R\sigma}^{r} & G_{d_{2}d_{1}\sigma}^{r} &
G_{d_{2}d_{2}\sigma}^{r}%
\end{array}
\right]  .
\end{equation}
The bare Green functions $g_{\alpha}^{r}$ in the leads with wide-band
approximation are taken in the form $g_{\alpha}^{r}=-i\pi\rho$, and
$g_{d_{i}d_{i}\sigma}^{r}$ on the dots can be obtained exactly by the equation
of motion method%
\begin{equation}
g_{d_{i}d_{i}\sigma}^{r}\left(  \omega\right)  =\frac{\omega-\epsilon
_{i}-U+Un_{i-\sigma}}{\left(  \omega-\epsilon_{i}\right)  \left(
\omega-\epsilon_{i}-U\right)  },
\end{equation}
where $\rho$ is the density of states of the leads and $n_{i-\sigma}$ is the
occupation number with spin $-\sigma$ on the $i$th dot. The self-energy
neglecting higher-order terms is written as the tunneling matrix%
\begin{equation}
\Sigma_{\sigma}^{r}\left(  \omega\right)  \equiv\left[
\begin{array}
[c]{cccc}%
0 & t_{LR} & t_{Ld_{1}} & 0\\
t_{LR}^{\ast} & 0 & 0 & \widetilde{t}_{Rd_{2}}\\
t_{Ld_{1}}^{\ast} & 0 & 0 & t_{d_{1}d_{2}}\\
0 & \widetilde{t}_{Rd_{2}}^{\ast} & t_{d_{1}d_{2}}^{\ast} & 0
\end{array}
\right]  ,
\end{equation}
with $\widetilde{t}_{Rd_{2}}=t_{Rd_{2}}e^{-i\sigma\varphi}$. This
approximation is sufficient above the Kondo temperature to study the Fano
resonance and spin accumulation. The occupation number in Eq. (10) is
determined self-consistently by the equation%
\begin{equation}
n_{i\sigma}=-i\int\frac{d\omega}{2\pi}G_{d_{i}d_{i}\sigma}^{<}\left(
\omega\right)  .
\end{equation}

\section{Differential conductance}

In numerical calculation we set $\rho=1$, and all the tunneling matrix
elements are simplified as constants $t_{LR}=0.1$, $t_{Ld_{1}}=t_{Rd_{2}}%
=0.4$, and $t_{d_{1}d_{2}}=0.8$. The temperature is set to $k_{B}T=0.0001$
($T\approx1.16$ $%
%TCIMACRO{\unit{K}}%
%BeginExpansion
\operatorname{K}%
%EndExpansion
$). The Fermi level is here set as the origin of energy, and the energies are
in units of $\Gamma\equiv2\pi\rho t_{\alpha d_{i}}^{2}\approx1$. A symmetric
bias $V$ applied to two leads results in a chemical potential difference so
that $\mu_{L}=-\mu_{R}=V/2$. We have determined a distinguishing point in the
plot of differential conductance versus bias voltage, as shown later in Fig.
5(a). The value of bias at this point is about $V=0.2$ which equals ten
percent of Coulomb repulsion strength $U=2$. Below this value differential
conductance differs from zero bias conductance within one percent and is
basically considered as a constant. Above this value differential conductance
departs from the constant and noticeably changes with both bias and RSO
coupling strength. Therefore, we discuss the system defined in weak and strong
bias conditions by regions below and above $V=0.2$, respectively.

\subsection{Weak bias}

Excluding the cases with level position on dot well below and above the
equilibrium chemical potential, an electron in lead with weak bias potential
can hardly overcome the Coulomb energy to fill an occupied state. The
interference is thus expected to arise mainly from the mixing of electron
waves through a molecular state and the bridge channel.%

%TCIMACRO{\FRAME{fhFU}{4.1459in}{3.7118in}{0pt}{\Qcb{(Color online) Conductance
%spectrum with weak bias $V=0.02$ at (a) $\varphi=0$, (b) $\varphi=\pi/2$, and
%(c) $\varphi=\pi$. (d) Differential conductance with $\epsilon_{1}%
%=\epsilon_{2}\equiv\epsilon$ for $\varphi=0,\pi/2,\pi$.}}{}{fig2.ps}%
%{\special{ language "Scientific Word";  type "GRAPHIC";
%maintain-aspect-ratio TRUE;  display "USEDEF";  valid_file "F";
%width 4.1459in;  height 3.7118in;  depth 0pt;  original-width 6.2405in;
%original-height 5.5815in;  cropleft "0";  croptop "1";  cropright "1";
%cropbottom "0";  filename 'FIG2.ps';file-properties "XNPEU";}}}%
%BeginExpansion
\begin{figure}
[h]
\begin{center}
\includegraphics[
height=3.7118in,
width=4.1459in
]%
{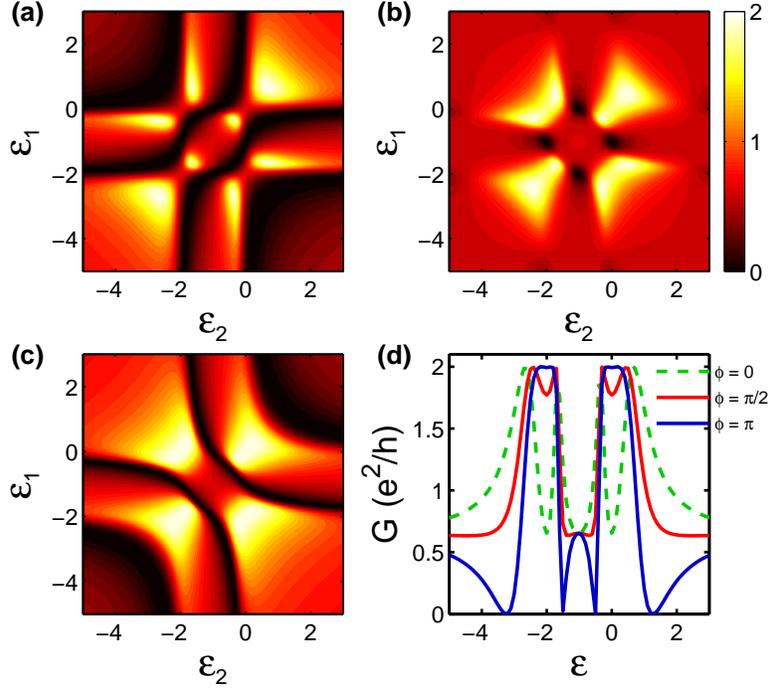}%
\caption{(Color online) Conductance spectrum with weak bias $V=0.02$ at (a)
$\varphi=0$, (b) $\varphi=\pi/2$, and (c) $\varphi=\pi$. (d) Differential
conductance with $\epsilon_{1}=\epsilon_{2}\equiv\epsilon$ for $\varphi
=0,\pi/2,\pi$.}%
\end{center}
\end{figure}
%EndExpansion
Figures 2(a)--(c) show the whole conductance spectrum as a function of
individual dot levels $\epsilon_{1}$\ and $\epsilon_{2}$ for three RSO
coupling strength $\varphi=\alpha_{R}m^{\ast}L/\hbar^{2}=0,\pi/2,\pi$. We can
clearly observe that the total conductance is strongly affected by the RSO
coupling, including the phases and the positions of the resonances and
antiresonances, which we then discuss. Consider the case $\epsilon
_{1}=\epsilon_{2}\equiv\epsilon$, as shown in Fig. 2(d). In the absence of RSO
coupling two Coulomb peaks are located at $\epsilon=0$ and $\epsilon=-U$. The
single-particle energy $\epsilon$ is modified by the interdot coupling and the
tunneling couplings of the dots to the leads. Each Coulomb peak thus split
into two peaks associated with the bonding and antibonding states. Without
bridge channel the electron transport is described by two pairs of
Breit-Wigner resonances, with a Coulomb gap in between, at the energies of
molecular states and unaffected by the RSO interaction. In present
configuration, a change in the phase difference between bonding and
antibonding resonances is here observed, when the RSO coupling is varied.
Electrons passing through two different pathways result in the quantum
interference, which is characterized by the Fano resonance corresponding to an
asymmetric line shape in the conductance curve. The Fano effect at $\varphi=0$
in Fig. 2(d) is modified since no antiresonance is found in the conductance.
This disappearance of antiresonances is ascribed to the strong overlap within
each pair of Fano resonances due to the large interdot coupling, which results
in the mergence of antiresonances.\cite{Joe05} As $\varphi$ increases, four
antiresonances occur at $\varphi=\pi$ as a consequence of destructive quantum
interference, and the value of the conductance remains the same at
$\epsilon=-1$. During the spin precession on dots, a spin-dependent phase
shift takes place in dot-lead matrix elements $t_{Rd_{2}}e^{-i\sigma\varphi}$.
From $\varphi=0$ to $\varphi=\pi$ the phase parameter $s$ of dot-lead matrix
element is altered from $s=+1$ to $s=-1$ in the complex plane for both up- and
down-spin electrons,\cite{Silva02} which leads to a phase evolution of
electron transport through four molecular states. Each pair of Fano resonances
at $\varphi=\pi$ is thus still out of phase within.\cite{Lee08} Moreover, one
can see that the RSO effect on the interference is present to offset the
suppression of the Fano antiresonances, because the typical Fano line shape is
completely restored.%

%TCIMACRO{\FRAME{fhFU}{4.152in}{1.9164in}{0pt}{\Qcb{(Color online) Differential
%conductance for $\epsilon_{2}=-3,-1,1$ at (a) $\varphi=0$ and (b) $\varphi
%=\pi$.}}{}{fig3.ps}{\special{ language "Scientific Word";  type "GRAPHIC";
%maintain-aspect-ratio TRUE;  display "USEDEF";  valid_file "F";
%width 4.152in;  height 1.9164in;  depth 0pt;  original-width 6.2483in;
%original-height 2.8617in;  cropleft "0";  croptop "1";  cropright "1";
%cropbottom "0";  filename 'FIG3.ps';file-properties "XNPEU";}}}%
%BeginExpansion
\begin{figure}
[h]
\begin{center}
\includegraphics[
height=1.9164in,
width=4.152in
]%
{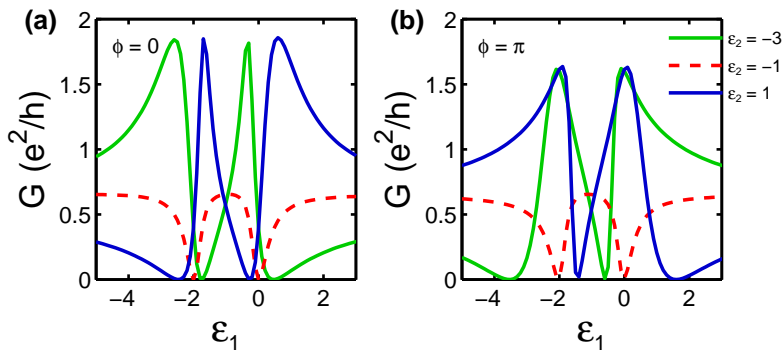}%
\caption{(Color online) Differential conductance for $\epsilon_{2}=-3,-1,1$ at
(a) $\varphi=0$ and (b) $\varphi=\pi$.}%
\end{center}
\end{figure}
%EndExpansion
We now address the electron transport through a tunable level, while the other
is fixed at particular energy position. Hence the conductance versus the dot
level $\epsilon_{1}$ is plotted at three values of $\epsilon_{2}=-3,-1,1$
under the RSO interaction. In Fig. 3(a), with $\varphi=0$, the conductance
shows the typical two-level Fano line shapes. The Fano parameter is
continuously altered by different values of $\epsilon_{2}$ from the left side
to the right side of $\epsilon_{2}=-1$, corresponding to negative value
through zero to positive value. In Fig. 3(b), with $\varphi=\pi$, the
conductance still shows the Fano line shapes. However, the Fano parameters
change sign except the curve at $\epsilon_{2}=-1$, in comparison with the
corresponding curves in Fig. 3(a). We note that in conformity with all
molecular states varied by the RSO coupling, as shown in Fig. 2(d), the phases
of two resonances are here simultaneously changed from the observation on the
same sign of Fano parameters.

\subsection{Strong bias}

In addition to the coherent transport, a strong bias shifts the energies of
the electrons in two leads. The nonequilibrium situation, with a bias voltage
$\mu_{L}=-\mu_{R}=2$ across the ring-dot system, is built up and contributes
to the electrical current through the device. With this increase of the bias
window excited states become accessible and participate in the transmission.
Meanwhile, the charge fluctuation can occur, and the interference between
molecular states is enlarged. The mixture of various interference among bridge
channel and molecular states thus complicates the transport through the device.%

%TCIMACRO{\FRAME{fhFU}{4.1459in}{3.7118in}{0pt}{\Qcb{(Color online) Conductance
%spectrum with strong bias $V=4$ at (a) $\varphi=0$, (b) $\varphi=\pi/2$, and
%(c) $\varphi=\pi$. (d) Differential conductance with $\epsilon_{1}%
%=\epsilon_{2}\equiv\epsilon$ for $\varphi=0,\pi/2,\pi$.}}{}{fig4.ps}%
%{\special{ language "Scientific Word";  type "GRAPHIC";
%maintain-aspect-ratio TRUE;  display "USEDEF";  valid_file "F";
%width 4.1459in;  height 3.7118in;  depth 0pt;  original-width 6.2405in;
%original-height 5.5815in;  cropleft "0";  croptop "1";  cropright "1";
%cropbottom "0";  filename 'FIG4.ps';file-properties "XNPEU";}}}%
%BeginExpansion
\begin{figure}
[h]
\begin{center}
\includegraphics[
height=3.7118in,
width=4.1459in
]%
{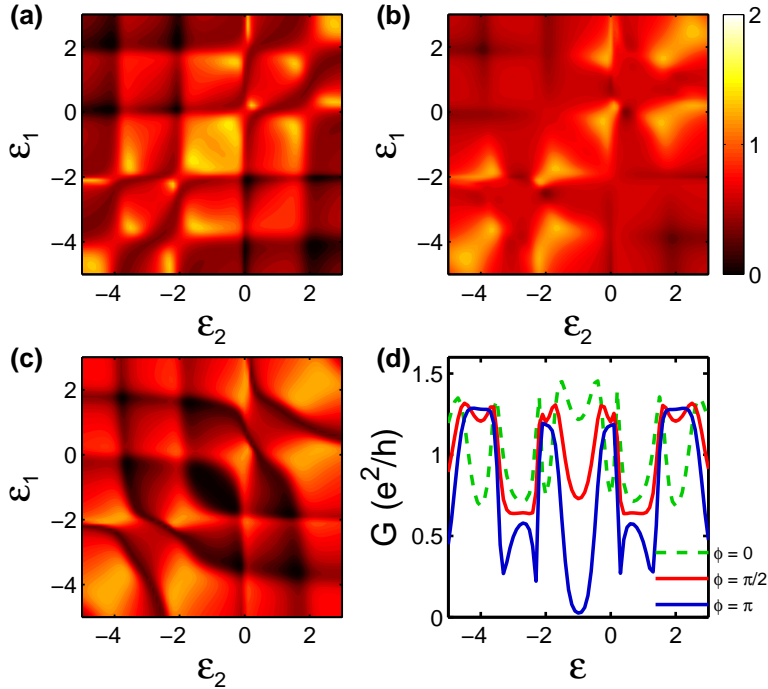}%
\caption{(Color online) Conductance spectrum with strong bias $V=4$ at (a)
$\varphi=0$, (b) $\varphi=\pi/2$, and (c) $\varphi=\pi$. (d) Differential
conductance with $\epsilon_{1}=\epsilon_{2}\equiv\epsilon$ for $\varphi
=0,\pi/2,\pi$.}%
\end{center}
\end{figure}
%EndExpansion
Figures 4(a)--(c) show the whole conductance spectrum as a function of
individual dot levels $\epsilon_{1}$\ and $\epsilon_{2}$ for three RSO
coupling strength $\varphi=0,\pi/2,\pi$. We can clearly observe that the total
conductance here is still affected by the RSO coupling and is very different
from that in the weak bias condition. Consider the case $\epsilon_{1}%
=\epsilon_{2}\equiv\epsilon$, as shown in Fig. 4(d). In the absence of RSO
coupling there are eight peaks in the conductance curve. These peaks are
distinctly located around $\epsilon=-4,-2,0,2$. In each pair of peaks the
energy separation is associated with the interdot coupling and the tunneling
couplings of the dots to the leads. The doubled peaks are ascribed to the
electron conduction when each molecular state is aligned with either of leads.
One can see that the maximum values of these peaks are lower than those in the
weak bias condition. With increasing $\varphi$ the splits within pairs of
peaks decrease and the antiresonances in between form. This transition is the
same as that in Fig. 2(d) due to the $s$-dependence of couplings to the lead,
except for the suppressed Fano antiresonances at $\varphi=\pi$. The change in
Fano line shapes indicates that the phase of the electron transmitting via
both bonding and antibonding states is shifted with difference $\pi$ by the
RSO coupling. In addition, the value of conductance at $\epsilon=-1$, unlike
weak bias, does not remain the same. We thus study further the behavior of the
conductance at this point, as shown in Fig. 5(a). In weak bias condition,
$V<0.1U$, the conductance is almost unaffected by the RSO coupling. However,
the conductance with different values of RSO coupling strength can be enhanced
to two or be reduced to zero at certain values of larger bias voltage.%

%TCIMACRO{\FRAME{fhFU}{4.1502in}{3.9902in}{0pt}{\Qcb{(Color online) (a)
%Differential conductance at $\epsilon_{1}=\epsilon_{2}\equiv\epsilon=-1$ for
%$\varphi=0,\pi/4,\pi/2,3\pi/4,\pi$. Differential conductance for $\epsilon
%_{2}=-3,-1,1$ at (b) $\varphi=0$, (c) $\varphi=\pi/2$, and (d) $\varphi=\pi$%
%.}}{}{fig5.ps}{\special{ language "Scientific Word";  type "GRAPHIC";
%maintain-aspect-ratio TRUE;  display "USEDEF";  valid_file "F";
%width 4.1502in;  height 3.9902in;  depth 0pt;  original-width 5.8072in;
%original-height 5.5815in;  cropleft "0";  croptop "1";  cropright "1";
%cropbottom "0";  filename 'FIG5.ps';file-properties "XNPEU";}}}%
%BeginExpansion
\begin{figure}
[h]
\begin{center}
\includegraphics[
height=3.9902in,
width=4.1502in
]%
{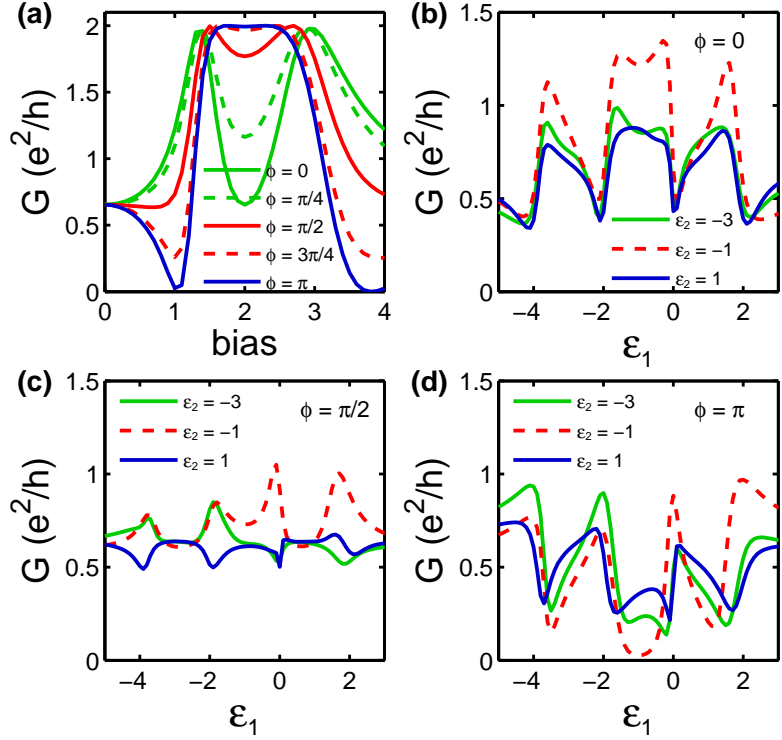}%
\caption{(Color online) (a) Differential conductance at $\epsilon_{1}%
=\epsilon_{2}\equiv\epsilon=-1$ for $\varphi=0,\pi/4,\pi/2,3\pi/4,\pi$.
Differential conductance for $\epsilon_{2}=-3,-1,1$ at (b) $\varphi=0$, (c)
$\varphi=\pi/2$, and (d) $\varphi=\pi$.}%
\end{center}
\end{figure}
%EndExpansion
Consider now the situation with a fixed level $\epsilon_{2}=-3,-1,1$. Figures
5(b)--(d) show the conductance as a function of the dot level $\epsilon_{1}$
for $\varphi=0,\pi/2,\pi$. At $\varphi=0$, as shown in Fig. 5(b), the sign
change of the Fano parameter, symmetric with respect to the curve at
$\epsilon_{2}=-1$ in weak bias condition, vanishes. Instead, the three curves
behave similarly. Each curve is composed of two pairs of strongly deformed
two-level Fano line shapes with opposite sign of the Fano parameter.
Resonances above and below $\epsilon_{1}=-1$ are in phase respectively and out
of phase mutually. This feature of conductance curves shows that the phases of
resonances are no longer strongly altered by the level position $\epsilon_{2}$
but $\epsilon_{1}$ above or below $\epsilon_{1}=-1$. In Fig. 5(d), with
$\varphi=\pi$, the phases of four resonances are simultaneously changed by the
RSO coupling, including the curve at $\epsilon_{2}=-1$. Therefore, peaks and
valleys in Fig. 5(a) are now switched.

\section{Spin polarization}

In the absence of external magnetic field threading through the dots, a
spin-unpolarized DQD is expected, with occupation numbers decreasing with
increasing level position. An electric field established on semiconductor
heterostructure serves as an effective magnetic field on moving electrons in
2DEG, and in consequence leads to a spin splitting. This spin splitting
between up- and down-spin electrons combined with quantum interference can
bring the spin-polarized occupation. In the preceding analysis we show that
the spin accumulation on DQD can be induced in this device and depends on the
variation in RSO coupling strength, meaning that the electron spin can be
manipulated by simply tuning a gate voltage.%

%TCIMACRO{\FRAME{fhFU}{4.1286in}{3.813in}{0pt}{\Qcb{(Color online) (a) Up- and
%down-spin electron occupation on the 2nd dot with $\epsilon_{1}=\epsilon
%_{2}\equiv\epsilon$ at $\varphi=\pi/2$. Green curves are for $V=1$ and red
%curves are for $V=4$. (b) Spin polarization at $\varphi=\pi/8$ for
%$V=0.02,1,2,4$. Spin polarization for $\varphi=\pi/8,\pi/4,\pi/2,3\pi/2$ at
%(c) $V=1$ and (d) $V=4$.}}{}{fig6.ps}{\special{ language "Scientific Word";
%type "GRAPHIC";  maintain-aspect-ratio TRUE;  display "USEDEF";
%valid_file "F";  width 4.1286in;  height 3.813in;  depth 0pt;
%original-width 6.0303in;  original-height 5.5668in;  cropleft "0";
%croptop "1";  cropright "1";  cropbottom "0";
%filename 'FIG6.ps';file-properties "XNPEU";}}}%
%BeginExpansion
\begin{figure}
[h]
\begin{center}
\includegraphics[
height=3.813in,
width=4.1286in
]%
{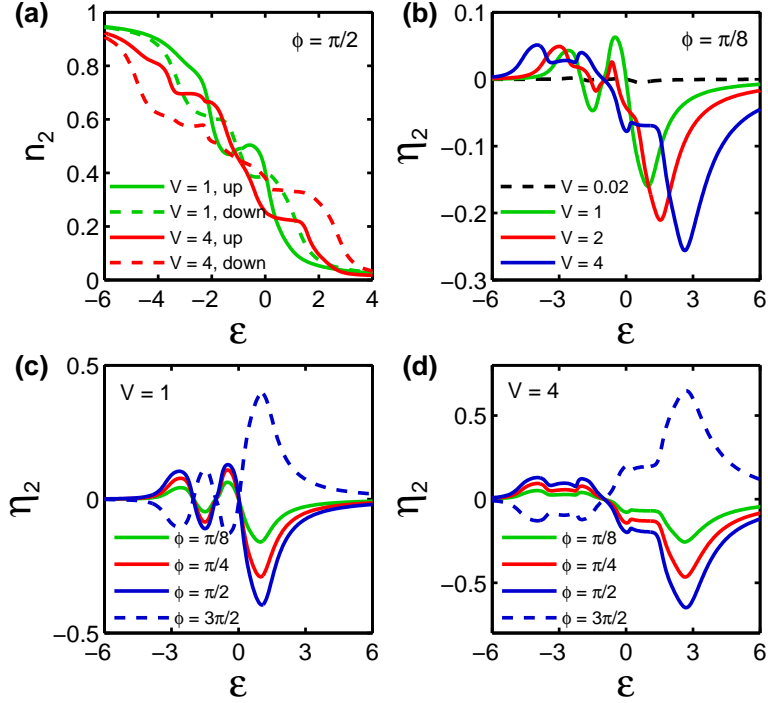}%
\caption{(Color online) (a) Up- and down-spin electron occupation on the 2nd
dot with $\epsilon_{1}=\epsilon_{2}\equiv\epsilon$ at $\varphi=\pi/2$. Green
curves are for $V=1$ and red curves are for $V=4$. (b) Spin polarization at
$\varphi=\pi/8$ for $V=0.02,1,2,4$. Spin polarization for $\varphi=\pi
/8,\pi/4,\pi/2,3\pi/2$ at (c) $V=1$ and (d) $V=4$.}%
\end{center}
\end{figure}
%EndExpansion
Figure 6(a) shows the electron occupation on the 2nd dot as a function of a
common gate voltage $\epsilon_{1}=\epsilon_{2}\equiv\epsilon$ at RSO coupling
strength $\varphi=\pi/2$. In the presence of RSO interaction the spin
degeneracy is lifted, and the dot is spin-polarized with spin accumulation
$n_{2\uparrow}-n_{2\downarrow}$ changing with the level position. At bias
$V=4$, the curves for up and down spins intersect at $\epsilon=-1$. The spin
accumulation on the left side of this point is positive while on the other
side is negative. In addition to this crossing point, two more points around
$\epsilon=0$ and $\epsilon=-2$ in the curves at $V=1$ are observed. By these
well-separated curves for up and down electron spins, a spin-polarized dot is
here verified. We then study the influence of tunable bias and RSO interaction
on the SP, defined as $\eta_{i}\equiv\left(  n_{i\uparrow}-n_{i\downarrow
}\right)  /\left(  n_{i\uparrow}+n_{i\downarrow}\right)  $.

Let us first consider the situation, at fixed RSO coupling strength
$\varphi=\pi/8$, with different values of bias from weak to strong condition,
as shown in Fig. 6(b). SP on the DQD is tiny, $<5\%$, when weak bias is
applied to the device. As the value moves to strong bias region, the SP grows
over $10\%$ and obviously varies with $\epsilon$. In the curves at $V=1$ and
$V=2$, there are three crossing points in negative $\epsilon$ region. We can
clearly see that two crossing points, located at $\epsilon=0,-2$ when $V=1$,
move toward each other as the bias increases. Subsequently they collide with
the central one to disappear at about $V=2.3$, leaving single crossing point
for larger value of bias $V=4$, for example. Thus, in strong bias condition,
SP invariably changing sign at $\epsilon=-1$ is observed. This characteristic
of SP meanwhile provides the possibility for the operation of spin flipping on
the dot. In positive $\epsilon$ region a valley exists in each curve. This
valley shifts to larger value of $\epsilon$ and becomes deeper with increasing
bias. With such small RSO coupling strength, the maximum value of SP is found
to be $-25\%$. For larger value $\varphi=\pi/2$, the SP can even reach $-65\%$
at $V=4$, as shown later in Fig. 6(d).

Now we discuss the behavior of induced SP altered by the RSO coupling
strength. Figure 6(c) shows the SP, with $V=1$, as a function of $\epsilon$
for $\varphi=\pi/8,\pi/4,\pi/2,3\pi/2$. In negative $\epsilon$ region one can
observe that the SP fluctuates about zero, which is consistent with the
corresponding occupation in Fig. 6(a). With increase of RSO coupling the SP
increases as well, with profile of the curve unchanged. The maximum values of
the fluctuation and the valley go from $4\%$ to $10\%$ and from $-16\%$ to
$-40\%$, respectively. Once we flip $\varphi$ with respect to $\pi$, from
$\pi/2$ to $3\pi/2$, we can see clearly that the SP totally flips with respect
to zero. Hence the peak and the valley are switched. In Grundler's
experiment,\cite{Grundler00} with a 2DEG on InAs-based heterostructure, the
measured maximum value of Rashba parameter is $\alpha_{R}\approx0.4$ $%
%TCIMACRO{\unit{eV}}%
%BeginExpansion
\operatorname{eV}%
%EndExpansion%
%TCIMACRO{\unit{\U{212b}}}%
%BeginExpansion
\operatorname{\text{\AA}}%
%EndExpansion
$. By this value of $\alpha_{R}$, $\varphi=\pi/4$ corresponds to typical
self-assembled dot size $L/2\approx20$ $%
%TCIMACRO{\unit{nm}}%
%BeginExpansion
\operatorname{nm}%
%EndExpansion
$, while $\varphi$ can reach $5.67$ with conventional gate-defined DQD size
$L=300$ $%
%TCIMACRO{\unit{nm}}%
%BeginExpansion
\operatorname{nm}%
%EndExpansion
$. Therefore, both the magnitude and the flip of SP, as observed here in Figs.
6(b) and (c), can be realized experimentally. Concerning the case $V=4$, as
shown in Fig. 6(d). Two characteristics of SP, i.e. increase and reversal with
varied $\varphi$, exist similarly as the case $V=1$ except that the
fluctuations vanish and the overall values of SP are higher.%

%TCIMACRO{\FRAME{fhFU}{4.1087in}{1.9666in}{0pt}{\Qcb{(Color online) Occupation
%difference of up spin (a) at $V=1$ for $\varphi=\pi/8,\pi/4,\pi/2$ and (b) at
%$\varphi=\pi/8$ for $V=1,2,4$.}}{}{fig7.ps}%
%{\special{ language "Scientific Word";  type "GRAPHIC";
%maintain-aspect-ratio TRUE;  display "USEDEF";  valid_file "F";
%width 4.1087in;  height 1.9666in;  depth 0pt;  original-width 6.0027in;
%original-height 2.8522in;  cropleft "0";  croptop "1";  cropright "1";
%cropbottom "0";  filename 'FIG7.ps';file-properties "XNPEU";}}}%
%BeginExpansion
\begin{figure}
[h]
\begin{center}
\includegraphics[
height=1.9666in,
width=4.1087in
]%
{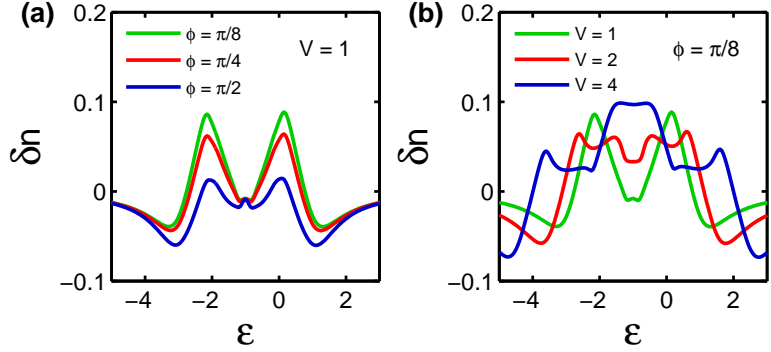}%
\caption{(Color online) Occupation difference of up spin (a) at $V=1$ for
$\varphi=\pi/8,\pi/4,\pi/2$ and (b) at $\varphi=\pi/8$ for $V=1,2,4$.}%
\end{center}
\end{figure}
%EndExpansion
At last we address the relation between electron occupation on two dots. The
difference of up spin, $\delta n\equiv n_{2}-n_{1}$, as a function of
$\epsilon$ is plotted at $\varphi=\pi/8,\pi/4,\pi/2$ and $V=1,2,4$, as shown
in Figs. 7(a) and (b), respectively. In both figures $\delta n$ is highly
symmetric with respect to $\epsilon=-1$. The increase of RSO coupling appears
to suppress the difference. Increasing bias, moreover, shifts $\delta n$
toward the negative and causes the peak splitting, which is ascribed to more
Coulomb peaks entering the transport window.

\section{Conclusion}

We have studied the quantum transport through a ring with embedded DQD, in
which RSO interaction and on-site Coulomb repulsion are considered. Our
analysis focuses on quantum interference and spin accumulation, with influence
of both RSO coupling and nonequilibrium effect. For $V<0.1U$, conductance in
this linear response regime shows two pairs of Fano resonances. With gradual
increase of $\varphi=0$ to $\pi$, antiresonances are well resolved, because
dot-lead matrix elements for up- and down-spin electrons are altered by the
spin-dependent phase associated with spin precession on dots. Consequently
both bonding and antibonding resonances are varied with phase difference $\pi
$. For $V>0.1U$, differential conductance noticeably changes with bias and RSO
coupling strength. Such large bias window leads to more conductance peaks,
when each molecular state is aligned with either of leads. However, the
strength of resonances is notably suppressed due to the mixture of various
kind of interference, arising from accessible excited states. In the study of
transport through the tunable $\epsilon_{1}$, similar Fano line shapes in
conductance for fixed $\epsilon_{2}=-3,-1,1$ show that the dependence of
corresponding phases on $\epsilon_{2}$ is strongly weakened. Instead, the
dependence on $\epsilon_{1}$ above or below $\epsilon_{1}=-1$ is observed.
Furthermore, with the help of a bridge channel, SP on serially coupled DQD can
be achieved through the RSO interaction. We note that both of the increase of
bias and RSO coupling strength can strengthen this SP. As $V<2.3$, SP behaves
similarly, with a fluctuation about zero for $\epsilon<0$ and a deep valley
for $\epsilon>0$, except that the deeper valley shifts to higher energy
position as bias increases. As $V>2.3$, the fluctuation is removed, leaving a
substantial sign-changed point at $\epsilon=-1$. Meanwhile, SP can
instantaneously flip once $\varphi$ is tuned to cross $\pi$, which leads to a
deep valley reversed to a high peak. Therefore, we have shown that the SP can
be controlled to point along either up or down by simply switching the dot
levels and the RSO coupling strength with respect to $\epsilon=-1$ and
$\varphi=\pi$, respectively.

\section{Acknowledgments}

This work was supported by the Republic of China National Science Council
under\ Grant No. 95-2112-M-002-044-MY3.

\bibliographystyle{apsrev}
\bibliography{acompat,kwchendb}

\end{document}